\documentclass[12pt,epsfig]{article}
\usepackage{hyperref}

\begin{document}

\begin{flushright}
SLAC-PUB-8357\\
arXiv:physics/0007040\\
February 2000
\end{flushright}

\begin{center}
{\large \bf{Physicists Thriving with Paperless Publishing} }

\bigskip\bigskip

Heath B. O'Connell\footnote{Work 
supported by the US
         Department of Energy contract DE-AC03-76SF00515.\\
{\it Invited talk at the AAAS Annual Meeting and Science Innovation
Exposition, February 17-22, 2000 in Washington, DC}}\\
Stanford Linear Accelerator Center\\ 
Stanford CA 94309\\
{\tt hoc@slac.stanford.edu}\\
{\tt http://www.slac.stanford.edu/grp/lib/people/hoconnell.html}
\end{center}

The Stanford Linear Accelerator Center (SLAC) and Deutsches Elektronen
Synchrotron (DESY) libraries have been comprehensively cataloguing the
High Energy Particle Physics (HEP) literature online since 1974. The core
database, SPIRES-HEP, now indexes over 400,000 research articles, with
almost 50\% linked to fulltext electronic versions (this site now has over
15 000 search hits per day). This database motivated the creation of the
first site in the United States for the World--Wide Web at SLAC. With this
database and the invention of the Los Alamos E-print archives in 1991, the
HEP community pioneered the trend to ``paperless publishing" and the trend
to paperless access; in other words, the ``virtual library."  We examine
the impact this has had both on the way scientists research and on
paper-based publishing. The standard of work archived at Los Alamos is
very high. 70\% of papers are eventually published in journals and another
20\% are in conference proceedings. As a service to authors, the
SPIRES-HEP collaboration has been ensuring that as much information as
possible is included with each bibliographic entry for a paper. Such
meta-data can include tables of the experimental data that researchers can
easily use to perform their own analyses as well as detailed descriptions
of the experiment, citation tracking, and links to full-text documents.

\newpage

\section{Introduction}

The World Wide Web is now ubiquitous. At a time when every advertisement
on television seems to end with a URL, it is of some interest to note that
the SLAC Library had the first Web server in North America (indeed the
first outside Europe), which it set up to aid the research of High
Energy/Particle physicists. In this talk we explore the role technology
has played in the organisation and dissemination of information in High
Energy Physics (HEP), where paper is now a medium of last resort, and
discuss why HEP provided the ideal conditions for the rapid adoption of
new technology.

\section{The Compulsive Communicators}

Communication has long played a vital role in physicists' research.  Not
content to wait for the publication of their work in journals, they, or
perhaps their departments, would send out {\it preprints} of their work to
institutions or other researchers they personally knew (once the
duplication process became affordable). Naturally because of the cost and
effort involved in this, large departments and famous scientists enjoyed a
numerical advantage both in sending and receiving. However, for every poor
soul in a small, out of the way department lamenting intellectual
starvation, a well known researcher would be suffering from an overstuffed
mailbox. By the 1960's the sheer number of preprints had somewhat
perversely made the communication of research {\it more} difficult, due to
information overload.

Since its foundation in 1962, the SLAC Library, at the request of the
director, W.K.H.~Panofsky, had been actively collecting new preprints (as
a world centre for physics, it attracted some three thousand articles per
year).  The details of each preprint were entered into a card catalogue,
including every author, no matter how many there were on the paper, which
for an experiment could easily be in the hundreds (the current record is
over 1000).  {From} this catalogue, the library published a weekly list of
new preprints for the SLAC community.

In 1967 work began at Stanford University on a computer database system
that would be able to handle a practically limitless number of large
bibliographic records.  It was decided early on that the SLAC Library's
preprint card catalogue would be a suitable test subject for this
database, and so in March 1968 the SLAC Library began participating in the
development and testing of SPIRES, the Stanford Physics Information
REtrieval System. SPIRES (later renamed the Stanford Public Information
REtrieval System) was envisioned as a fully searchable database, and each
bibliographic record would contain the preprint's author list (with each
author's institution), title, report number, date.

The remaining bibliographic element, which turned out to be of vital
importance, was the paper's reference list, that is the list of earlier
papers the present paper has cited \cite{cites}.  Only references to
published journal articles would be included (as these were the most
likely to be accurate), and they would be stored by the journal, volume
and first page number. The importance of these reference lists was with
SPIRES, one could search through them for a given paper, and thus find
every preprint that {\em cited} the paper, thereby establishing (in some
way) the impact this paper had in the literature. Naturally, this is of
considerable interest to physicists \cite{topcites}.

However, this activity was not limited to the New World. In 1962, the
Deutsches Elektronen-Synchrotron (DESY) in Hamburg, Germany, had begun to
publish a list named ``High Energy Physics - An Index,'' (HEPI) of all the
research literature it received. Included in this list was both the
published literature and unpublished preprints (whereas SLAC concentrated
mainly on preprints), amounting to at that time some nine thousand
documents per year. Like SLAC, DESY provided the title, author and date,
but instead of references, HEPI staff physicists assigned to each paper up
to 23 keywords (or, more often, key phrases) chosen from a controlled
list.

Naturally it made sense for the SLAC and DESY teams to work together in
their coverage of the particle physics literature, and so preparations for
this began. By June 1969 specifications for the conversion of the DESY
data to SPIRES format were complete.

\section{SPIRES and the Internet 1969--1990}

In 1968 the SLAC librarians, Louise Addis, Bob Gex and Rita Taylor were
funded by the the US Atomic Energy Commission (precursor to today's
Department of Energy), with the Division of Particles and Fields (DPF) of
the American Physical Society as a sponsor, to publish a weekly preprint
list for the entire particle physics community \cite{LXA}. This list of
all preprints entered in the past week could be easily generated from the
newly commissioned SPIRES-HEP database. In January 1969 the first
Preprints in Particles and Fields (PPF) list was sent out to over a
thousand eager subscribers. Therefore, to get your work known all around
the world (or at least the title of your work seen) all you would have to
do would be send your preprint to SLAC, and similarly you could browse the
titles of all the week's preprints by just looking at one list.  This
established the idea of a central repository for preprints.

A sister ``Anti-preprints" list (PPA) recorded the preprints that had
since been published in journals. This allowed the original preprints to
be discarded, and proved quite popular with the journal editors, who were
then able to match references to preprints with the published article.

By 1974 both the SLAC and DESY libraries (and in later years other
collaborators \cite{collab}) were working on the single HEP database, and
comprehensively cataloguing preprints, and by extension, the published
literature, in high-energy physics.

This new PPF age, though, was not perfect. Receiving lists of new
preprints and journal articles is all very well, but after several years
one is left once again with a stack of paper. Trying to find a copy of a
particular article or information on a certain subject can be very
difficult. The Internet allowed a further development.

The problem of finding just the right preprints from this ever growing
body of research was not an issue for the SLAC community. The SPIRES
database allowed searches by date, author, institution, report number,
title, keywords (assigned by DESY) and a number of other fields. Simply by
logging in to their computer, SLAC physicists were able to search through
thousands of papers, and the paper PPF list would serve as a newspaper,
with a similar longevity.  With the rise of the Internet anyone in the
world could now access this service. A program, QSPIRES~\cite{qspires},
developed in 1985 \cite{Kreitz:1996} by SLAC database systems developer
George Crane, allowed physicists to search the SPIRES database using
E-mail and other bitnet message protocols (which actually had priority
over E-mail and hence were faster).  One would send off an E-mail query to
the database and a swift reply would follow.  This made an enormous
difference for people outside SLAC, especially those outside the U.S., who
had previously only been able to access SPIRES, through rare and specially
issued SLAC accounts. People were able to find all the papers by a
particular author or from a certain institution, or find out how many
citations their work had \cite{cites}. The PPF and PPA lists could also be
sent by E-mail.

\section{{\TeX} and the Single Archive}

By 1990, lists of new articles were being sent to physicists by E-mail and
people were searching for articles in the SPIRES database via E-mail.
Everything was modern and electronic, up to the point where you actually
wanted to read a paper. In this case, the reader would have to request
that an author send a copy in the mail. Thus, if you were overseas it could
take weeks to months before you actually saw the article;  this was
clearly unacceptable.

Around this time another technical innovation had sufficiently matured to
obviate this problem. High energy physics articles usually contain a lot
of mathematics, which makes them difficult to write with a standard set of
alpha-numeric characters.  In the late 1970's, Stanford computer scientist
Donald Knuth \cite{knuth} had begun work on a special typesetting program,
\TeX\ (pronounced ``tek'' since the ``x'' is actually a Greek $\chi$),
that could display mathematics beautifully from source containing no
special characters \cite{tex}.  After 1983 when the current version was
released, the language soon proved very popular with physicists and
mathematicians as it gave them complete control over the production of
their documents.\footnote{This paper has been written using a derivative
of \TeX}

The underlying ``\TeX-file'' was simply a normal, {\em portable} text
file, in which the mathematics was written using certain rules which could
then be processed {\em on any computer} into a 
viewable final output format, such as
PostScript (PS),
\cite{PS}, changing 
\begin{verbatim}
\int_0^1\frac{e^x}{2\pi} dx 
\end{verbatim}
to $$\int_0^1 \frac{e^x}{2\pi} dx. $$
The PS file can then be printed out and read. Therefore, to send a paper,
all you would have to do is send the tex-file via E-mail and the person at
the other end could process it and print it out; another step in the
communication process had entered the future. By 1990 the use of \TeX\
in the HEP community was almost universal.

With preprint list distribution, SPIRES database searches, requests for
papers, and even the transfer of these papers being done through E-mail,
all stages of the process were electronic and pretty much immediate \dots
except one. SLAC was still receiving the E-prints as ``hard-copy".  Some
authors had begun sending their preprints as \TeX\ files, but for some
researchers this would unfortunately lead to overstuffed electronic
mailboxes, echoing a problem of the Paper Days.  In 1991, Paul Ginsparg of
Los Alamos National Laboratory (LANL) decided to do something about this.  
If authors could send their papers electronically to a central repository,
with author-supplied bibliographic data, a list could be sent out of each
{\em day's} additions and the preprints could be obtained directly from
this archive \cite{arxiv}. Here, TeX proved a godsend, as it provided a
means of storing rather small files that researchers could request and
then process into the viewable PS files at their home institutions.  In
August 1991, the first paper was sent to the Los Alamos archive. Twenty
seven papers were sent that month\footnote{In August 1999, over 800
particle physics papers were sent to the LANL archive.} and physicists
began receiving daily E-mails of the new papers sent to LANL.

Soon there were over 2000 subscribers to this daily E-mail notification
and it gradually began to replace SLAC's PPF list (though not the SPIRES
database itself) as the HEP community's ``newspaper'' --- its most
immediate source of new research information.  Far from competing, though,
the two services at LANL and SLAC, complemented each other nicely. SLAC
continued to process hard copy preprints that were never sent to LANL and
journal articles that had never been preprints, adding such information as
their references, experiment numbers, authors' institutions, etc.  The
electronic preprints, {\it E-prints}, began to account for a good deal of
the HEP literature, and Ginsparg's system allowed for considerable
automation and efficiency in SLAC's work. A good deal of useful
information could be harvested electronically from the files authors
supplied to LANL (though there was still much that had to be done by
hand).

This evolution from preprints to E-prints had another important facet. The
E-prints were assigned a unique number of the form {\it archive/0007040}
(eg 9501251 for the 251st paper in January 1995), which the SLAC Library
immediately began to store as a new element in each record in the SPIRES
database. In the past ``report numbers'' had been assigned to preprints by
the author's home institution in a less than completely systematic way and
with a variety of conventions, which made the tracking of citations too
difficult for the SPIRES database. Thus, as mentioned earlier, only
citations to published journal articles were recorded, and the many
citations made to an author's paper while it was an unpublished preprint
were lost.  The standardised E-print number changed all that, since these
numbers soon started appearing in reference lists. After some work, these
citations to unpublished works could be registered. It is worth noting
that the most cited HEP article in 1998 spent some ten months as an
unpublished E-print \cite{Maldacena:1998re}.  The next step was to ensure
that the database recognised that the published article and the E-print
were essentially the same paper, and thus, that the true number of
citations was the sum of the citations of both. In some instances this
lead to double counting, which needed special programming to eliminate, as
it became fashionable to include {\it both} the E-print number and the
journal reference when citing an article!

\section{The Mouse That Clicked}

With the SPIRES database and the LANL archive you could find a particular
paper and be reading it within minutes merely by sending E-mail, after
just a little work to process the \TeX\ file and print it out.  A lot had
changed in 20 years since the SLAC Library first started sending out a
list of all the preprints it had received. However, the guiding philosophy
behind computing is that no work a person has to do can ever be ``too
little.'' A new technology was about to make the research process even
easier.

In September 1991, SLAC physicist Paul Kunz, was visiting CERN, Geneva,
where he was shown the infant World Wide Web, which allowed information on
different computers to be accessed in a very user-friendly (and now
completely familiar) manner.  Realising the Web interface would be perfect
for searching the SPIRES database he told then librarian and SPIRES
database manager, Louise Addis, about it upon his return to SLAC. She was
quick to recognise the potential of this \cite{TBL} and by December SLAC
had the first WWW server in the Western Hemisphere.\footnote{There was no
Initial Public Offering.} The E-mail access-system became obsolete as more
and more physics departments installed the free software necessary to
reach the SPIRES WWW interface \cite{SPIRES}. A group was formed at SLAC,
the WWWizards \cite{wiz} to provide help with WWW technology
\cite{Deken:1998sm}.

The key feature of the WWW is linking. Therefore, when displaying search
results in the SPIRES database, a number of things could be linked to each
record. One of these was the full-text at Los Alamos which was being
stored in a minimal fashion as \TeX\ source. As mentioned before, \TeX\
files need to be processed before they can be printed out. In the early
stages, many of the source files contained errors or idiosyncrasies that
made them difficult to process. In order to enter all the relevant
information into the SPIRES database the SLAC Library staff needed 
to print out each paper, and so would process the \TeX\ files sent to Los
Alamos into PS files. At first this was done manually, which soon grew to
be very time consuming. Luckily, the SLAC Library was helped by Paul Mende
from Brown University who created an automatic procedure. Accompanying
figures could be faxed to the SLAC Library, where they would be converted
via NextFAX to a PS file. The PS files (of both text and figures) were
then stored on a server at SLAC. The record in the SPIRES database would
link to both the Los Alamos entry, where the \TeX\ source file was stored,
and the SLAC PS server, where the user could view the PS file.

Around the same time a similar script to convert \TeX\ source to PS had
been written at Los Alamos by Tanmoy Bhattacharya, a physicist
specialising in the heavily computational subfield of Lattice Gauge
Theory. Initially for personal use, this PS generating code was built into
a powerful tool by Bhattacharya, Rob Hartill and Mark Doyle. By mid 1995,
the LANL archive was producing its own PS files thereby freeing the SLAC
Library from this task.  Ultimately PS generation was incorporated into
the LANL submission process, and it came with an ultimatum: if your paper
couldn't be processed, it wouldn't be accepted.  Many authors were shocked
to find that the \TeX\ files they had previously sent to Los Alamos could
not be easily processed, thus making their work less accessible, and
welcomed this checking procedure. With the ability to generate PS from
directly from \TeX\ source, PS that had been generated from \TeX\ was no
longer acceptable, and by late 1996 submissions of such would be rejected
\cite{whyTex}. For whatever reasons, this led to some grumbling, but faced
with being denied the opportunity to post to LANL, there was no real
alternative to complying.\footnote{An interesting and oftentimes amusing
timeline of the development of the Los Alamos archive can be found at
\cite{LANLtime}.}

With all this in place the SPIRES database was linking to reliable PS
files generated and stored at the Los Alamos archive. The LANL archive
would also link to both references and citations stored in the SPIRES
database. This formed perhaps the first publically available full-text,
cross-linked scientific database. Paper had been eliminated from the
process and the virtual library was born \cite{Kreitz:1996}.

\section{The Many Hands Interpretation}

The model of 
authors sending their electronic texts to LANL has pushed the effort of
publication onto the researchers themselves, who benefit from an
inexpensive, immediate, wide-scale dissemination of their work. This idea
has led many computer scientists to ponder automated systems for indexing
and retrieving these full-text papers. How much of the work of collecting
information on the literature can be realistically facilitated by the
authors? The combination of the Web, the SLAC Library's automated systems
and the LANL archives provides an interesting testing ground for this
question.

The goal of the LANL archive is to be as automated as possible, so that it
can exist without administrative intervention (as opposed to the SPIRES
database). To this end it has a number of checks to ensure that all
submissions meet the entry requirements (one we have already discussed
ensures the \TeX\ file is successfully processed into PS).  The basic
bibliographic information (author, title, etc.) data is supplied by the
users in a neatly structured format that can be downloaded into the SPIRES
database automatically. This raw information requires only minor attention
from the Library staff who, among other things, ensure spelling
consistencies, and add in the author affiliations from the INSTITUTIONS
database.

Far and away the most time consuming part of this is collecting the
references of each paper, from which the citation searching is built
\cite{cites}.  As citation results reflect some degree of professional
accomplishment, physicists tend to be rather interested in them and a good
deal of their correspondence with the SLAC Library concerns omissions or
mistakes in the reference lists (which is exacerbated by the posting to
LANL of revised versions of E-prints, which often contain additional
references).  Originally these reference lists were typed into the
database, but as another happy spin-off of \TeX\ about 90\% of them can
now be extracted from the author's file.  Unfortunately, this only makes a
huge job large, as they still have to be checked by the Library staff, and
authors regularly confound the reference extracting program by adopting
imaginative new ways of writing the journal-volume-page sequence, or
simply making errors in a reference.

Obviously the LANL model suggested there should be some way to place the
burden of constructing and checking the reference list on the authors
themselves (as they do with bibliographic information), before they send
their paper out to the world.  Two things really stood in the way of this.
First, the original program that extracted the references was not
definitive enough for this sort of user-side checking, as there was no
real way of specifying which references the code would extract, and which
it would not.  The second problem was that with physicists being rather
busy, any new system would have to result in less, not more, work for it
to be widely adopted.

Once again, \TeX\ helped with the solution. The SPIRES team developed and
offered a new service: the database would display records in \TeX\ format,
with the exact information needed to construct the reference as an
additional tag. The authors then had a very simple process of cut and
paste to create their reference lists, and the tag would sit invisibly in
the \TeX\ file (not appearing in the final PS file) waiting to be
extracted at the SLAC library. The strict structure of the CITATION tag
and the simplicity of its extraction allowed for the creation of a
checking program that authors could use, \label{check} for example:
\begin{verbatim} 
\bibitem{O'Connell:1997} 
H.~B.~O'Connell, 
``Recent developments in $\rho\!-\!omega$ mixing,'' 
Austral.\ J.\ Phys.\ {\bf 50}, 255 (1997) 
[hep-ph/9604375]. 
%%CITATION = HEP-PH 9604375;%% 
\end{verbatim}
which becomes:\\

\noindent
H.~B.~O'Connell,
``Recent developments in $\rho\!-\!\omega$ mixing,''
Austral.\ J.\ Phys.\  {\bf 50}, 255 (1997)
[hep-ph/9604375].
\\

\noindent
So far over five hundred papers have been written using this system.

A second way we've been experimenting with author--supplied data is asking
them to help us with another thing that is of great interest to them:  
missing papers in our database. Traditionally adding them in was another
time consuming exercise that would require Library staff to get the
journal from the shelves and type in all the relevant information.  Using
a web form that authors can fill out themselves allows a paper to be added
to the database simply by cutting and pasting on the part of the Library
staff.

By inviting our users to help us in maintaining the database and
automating as many processes as possible, we have been able to accomplish
a lot of additional work without spending significant amounts of extra
time. This would not be possible, though, without the eagerness and
attention to detail that characterises the Physics community.

\section{All the news that's fit to link}

The WWW created a unique opportunity for the SPIRES database by providing
a system that would conveniently attach to any record all the related
information both inside the database and around the world in a very
compact manner. Over the past six years we have worked on making this
process as efficient as possible both in terms of computer programming and
the output display our users see \cite{galic}. I shall now discuss how
this works.

A single record in the Literature database might have certain {\it basic}
elements such as
\begin{itemize}
\item Title
\item Author (and author's institution) or corporate author
\item Date
\item Publication note
\end{itemize}
but the article could also be {\em hyperlinked} to 
\begin{itemize}
\item the appropriate record in the EXPERIMENTS database
\item full text of article at the journal server or the E-print server
\item the references of the article
\item other articles that cite it
\item experimental data in the REACTIONS database,
\item the record for the conference at which the paper was given
\end{itemize}
allowing the researcher to selectively explore any particular related item.
  Thus  an original record only ten lines long becomes a gateway to
information stored around the globe.

The construction of these links requires special care and co-operation
with outside services.  One particular case that followed an evolutionary
process was linking to the published version of the document on journal
home pages.  Optimally, this URL would point to a unique ``abstract
page,'' rendered in HTML so that the link to the journal server can be
fast. Once there the user can be presented with all the Journal services
such as full text in a variety of formats (the most common being
PS or the newer Portable Document File (PDF)). {From} the point of view of
the publisher, this means that a URL has to be found for every article.  
{From} our point of view this URL should be calculable from the
information we already have about the record, as it then permits us to run
in the URL automatically.  Articles have always been cited through the
journal-volume-page (JVP) convention, so it made sense that a number of
major publishers, including the American Physical Society \cite{APS} and
Elsevier \cite{NH} adopted a URL scheme based on these three elements.  
Setting up such a system does present something of a technical challenge
for the journals, but is well worth it in terms of presenting the simplest
possible interface to the outside world and providing reliable access to
their wares.

The journals also link back into our database, mainly for the references
of the paper (though publishers have proven less eager to link to our
database record of the actual paper). In an effort which I hope will
become more widely adopted, we have worked closely with the American
Physical Society's {\it Physical Review} \cite{PRD} to share the
bibliographic data (including reference lists) and either update an
existing record in our database, or add a new record, when they publish new
articles.

Where possible we have also tried to link to other literature databases.
Some, like the CERN Library's database \cite{cern}, have a significant
overlap with our system (there are currently over fifty thousand 
two-way links between SPIRES and CERN). Others, such as Harvard's {\it
Astrophysics Data System} \cite{ads}and the American Mathematical
Society's {\it Mathematical Reviews} (AMSMR) \cite{msnet} cover a much
smaller set of papers in our database, but from a different perspective
and connect the paper to other academic disciplines.  Here, once again, we
need to work with the other databases to ensure maximal efficiency and
reliability in this linking.  Our links to the Harvard database use a JVP
scheme, while those to CERN and AMSMR databases use unique record keys.
SPIRES can be linked to either way. There is also the consideration that
the link should bring additional information, rather than just repeating
what record in the SPIRES database (or else why bother?). Therefore, we
apply these links selectively.

\section{So why HEP?}

Why did the High Energy Physics community provide such a fertile ground
for this particular aspect of the so-called ``Information Revolution''? We
have touched on the reasons throughout this talk, but it is useful,
perhaps, to summarise them to understand how this model might be more
widely adopted (or why it might {\em not} be).

The first thing is that it has not been a revolution so much as an
evolution. Particle physicists have always been compulsive communicators.
Generally free of commercial or governmental restrictions, the habit of
sending out advanced copies of work goes back to the days of carbon paper.
They also have a long standing tradition of international
collaboration, so any new advances in communication technology are eagerly
adopted, and the high technological literacy facilitates this. This is
especially true of the experimental community whose collaborations can
have hundreds of members all over the world and whose experiments depend
on high speed computer network connections. It is worth noting that the
two SLAC physicists most closely involved in spinning the web at SLAC,
Paul Kunz and Tony Johnson are experimentalists, though I should also
point out that Paul Ginsparg of Los Alamos is a theorist.

In this environment the sheer quantity of existing literature necessitated
creating the organisation begun by the SLAC Library in the late 1960's.  
Through using this PPF list service, physicists became accustomed to the
idea of a central ``clearing house" for preprints.  As the use of E-mail
spread through the Physics community, particularly the High Energy sector,
the SPIRES E-mail interface was introduced in the mid 80's and before long
was being used by thousands of researchers in over forty countries.  The
{\TeX} typesetting language, favoured by those needing to write complex
mathematics, allowed one to send simple ASCII files using E-mail, thus
giving rise to the electronic distribution of preprints, which was
centralised by Ginsparg's E-print archive.  The World Wide Web then laid
the ground for a highly powerful way to integrate the various facets of
research communication.

It should be acknowledged that these are rather special conditions. It is
instructive to note that in other scientific communities where {\TeX} is
less heavily used, the adoption of the LANL archives has been noticeably
slower, though in principle there is nothing to stop them (despite being
built with {\TeX} in mind, the archive is not dependent on it, and will
accept submissions in a variety of formats). The {\TeX} fluent mathematics
community, on the other hand, though quick to adopt the new system,
perhaps lacked the detailed computing knowledge to play a leading role.

For different research communities there are other obstacles. Faced with
the incredibly rapid and wholehearted adoption of this E-print scheme by
high energy physicists, the traditional journal publishers found it best
to not to offer any real resistance. They had nothing to fear, really, as
over seventy percent of HEP papers sent to Los Alamos are eventually
published in journals and another twenty percent appear in conference
proceedings.  In other fields, with newer archives, this is still
something of a touchy subject, such as in the biological and medical
research communities, as can be seen in the Biochemical Society's
response \cite{bio} to the NIH's proposal for a preprint server for the
life sciences \cite{nih} (further reflections on eprints and the life 
sciences can be found at \cite{beagle}).

\section{Conclusion}

In this paper I have described the evolution of paperless publishing in a
particular academic community and how special circumstances conspired to
make this happen faster and more comprehensively than in any other field.
In doing this I have hoped to convey the sense that this was indeed a
process of evolution rather than revolution.

It may be interesting to speculate what the future holds.  Will more and
more of what used to be called ``bibliographic data,'' become
author--supplied ``metadata'' which annotates each article with various
related information? Can this metadata be shared to create cross-database,
and cross-discipline, linking~\cite{SFX}? Our experience shows that if you
make the procedure simple enough, and offer an advantage to the authors
for doing it (such as the ability to check their reference list), 
they {\em will} use it. The challenge now
is to bring this to other scholarly communities \cite{openArchives}.

We have begun in the SPIRES-HEP database to exploit Web technology to
handle information beyond the traditional bibliographic record or what one
could obtain from a print version (such as the full text and reference
list). These other facets range from the paper's citation list, which is
constantly growing (if the author is fortunate), to the home-page of the
experimental collaboration that wrote the paper, to what information
researchers might really want out of a paper, in a form far more useful
than could ever be delivered in paper (such as a computer file of
experimental data or 3-D computer images created by the author).

The fluidity of this new, electronic means of publication also has
implications for the traditional research process. At what point does an
author ``give the final answer'' and the paper become ``set in stone''?
The LANL archive, has neatly addressed this issue in an appropriate
manner, by archiving each version of an E-print with a date stamp,
documenting any modifications in the paper. Once the article is
published in a journal, it is a finished piece of work, with any
subsequent alterations handled via errata. Other authors may
{\em comment} on the published paper, to which the author may {\em reply},
but the comment and reply are treated as new articles, and the original is
left unchanged.  One journal, however, has broken away from the
traditional model. {\it Living Reviews in Relativity} \cite{lr} exists
solely on the Web, and allows authors to constantly revise their articles.
In part this is due to the pedagogical nature of the journal; it seeks to
provide reviews that aid learning, rather than publish original research.

Well aware of the difficulties of predicting anything in the Internet
world, I have merely tried, in concluding, to state some of the current
trends in electronic publication. 

\begin{center}
\large{\bf Acknowledgements}
\end{center}
I would like to thank Louise Addis, Paul Ginsparg, Marek Karliner,
Pat Kreitz,
Michael Peskin, Hartmut Preissner and Ann Redfield
for reading this
manuscript and offering helpful comments and suggestions. 
I have also enjoyed conversations on the topics mentioned here with 
Richard Dominiak, Mark Doyle, Paul Ginsparg, 
Tony Johnson, John Jowett and Paul Kunz.

\end{document}